\documentclass[12pt]{article} 
\usepackage{fullpage,sak, epsfig} 
 
\catcode`@=11 
\def\@begintheorem#1#2{\trivlist \item[\hskip \labelsep{\bf #1\ #2.}]} 
\def\@opargbegintheorem#1#2#3{\trivlist 
      \item[\hskip \labelsep{\bf #1\ #2.\ (#3)}]} 
\catcode`@=12

\newtheorem{Theorem}{Theorem}[section] 
\newtheorem{Lemma}[Theorem]{Lemma} 
\newtheorem{Cor}[Theorem]{Corollary} 
\newtheorem{Def}[Theorem]{Definition} 
\newtheorem{Prop}[Theorem]{Proposition} 
\newtheorem{Conj}[Theorem]{Conjecture}

\def\QED{\hfill\fbox{\phantom{.}}} 
 
\newenvironment{proof}{\pagebreak[1]{\narrower\noindent 
{\bf Proof:\quad\nopagebreak}}}{\QED\medskip} 
 
\def\compclassfont#1{{\sf{#1}}}

\newcommand{\AC}{\compclassfont{AC}} 
\newcommand{\ACC}{\compclassfont{ACC}} 
\newcommand{\QAC}{\compclassfont{QAC}} 
 
\newcommand{\QACC}{\compclassfont{QACC}} 
\newcommand{\QTC}{\compclassfont{QTC}}

\newcommand{\MOD}{\compclassfont{MOD}}

\newcommand{\ket}[1]{{|{#1} \rangle}} 
\newcommand{\bra}[1]{{\langle {#1}|}}

\newlength{\g} 
\newcommand{\mc}[1]{{\mathcal {#1}}}

\begin{document} 
 
\title{Quantum Lower Bounds for Fanout} 
\author{M.~Fang\footnote{Computer Science Department, Boston University, Boston, MA 02215, $\{$heroes$|$homer$\}$@bu.edu}
 \and 
S.~Fenner\footnote{Dept.~of CS and Eng.,
University of South Carolina, Columbia, SC 29208,
$\{$fenner$|$zhang29$\}$@cse.sc.edu}
\and 
F.~Green\footnote{Dept.~of Math and CS, 
Clark University, Worcester, MA 01610, fgreen@black.clarku.edu}
 \and 
S.~Homer\footnotemark[1]
\and 
Y.~Zhang\footnotemark[2]
}
\maketitle 
 
 
\noindent{\bf Abstract}: {We prove several new lower bounds for constant 
depth quantum circuits. The main result is that parity (and hence 
fanout) requires log depth circuits, when the circuits are composed 
of single qubit and arbitrary size Toffoli gates, and when they 
use only constantly many ancill\ae. Under this constraint, this 
bound is close to optimal. In the case of a non-constant number 
$a$ of ancillae, we give a tradeoff between $a$ and the required 
depth, that results in a non-trivial lower bound for fanout when 
$a = n^{1-o(1)}$.}

\section{Introduction}\label{sec:Intro} 
There has been significant recent progress in understanding the 
power of constant depth quantum circuits. Such circuits are of 
considerable interest as the first quantum  circuits will 
certainly be small circuits with limited gates and constant depth. 
Much of the progress in this area has been in showing that 
constant depth circuits are more powerful than their classical 
counterparts. However, these and other upper bounds seem to 
require the presence of a (reversible) quantum {\it fanout} gate. 
A fanout gate takes an arbitrary number of bits and fans out one 
of them by taking its XOR with each of the others. Here we 
consider the question of whether fanout gates are necessary for 
these upper bounds. We prove several lower bounds showing that 
fanout cannot be computed using only generalized (i.e., unbounded 
size) Toffoli and single qubit gates when the number of extra work 
bits (ancill\ae) that the circuit uses is limited. 
 
Fanout gates have proved to be unexpectedly powerful. 
Moore~\cite{moore99} first observed that fanout gates and parity gates, 
in the presence of single qubit gates using 0 ancillae, are 
equivalent up to depth 3. This was extended by Green et 
al.~\cite{GHMP}: fanout is even equivalent to {\it any} $\MOD_q$ 
function (for $q 
\ge 2$), which determines if the number of 1s in the input is not divisible 
by $q$. Here the equivalence is again up to constant depth, but 
using $O(n)$ ancill\ae. One may interpret this result by defining 
quantum circuit classes analogous to classical constant-depth 
circuit classes. For example, a reasonable analog of the classical 
unbounded fanin and fanout class $\AC^0$ is $\QAC^0_{wf}$, the 
class of constant depth quantum circuit families composed of 
single qubit, generalized Toffoli, and fanout gates. (Here the 
subscript ``{\it wf}\/" denotes ``with fanout.'') Similarly one 
may define quantum analogs of $\ACC(q)$ (called $\QACC(q)$) and 
$\ACC$ (called $\QACC$). Thus the equivalence of fanout with 
$\MOD_q$ implies that, for any $q > 2$, $\QAC^0_{wf} = \QACC(q) = 
\QACC$. Contrast this with the fact that $\AC^0 \not = \ACC$, and, 
for any distinct primes $q,p$, $\ACC(q) \not = \ACC(p)$~\cite{Razborov,
Smolensky}. More 
recently, H{\o}yer and Spalek~\cite{HS} have improved these 
results by proving these same $\QAC^0_{wf}$ circuits can compute 
threshold functions. Thus $\QAC^0_{wf} = \QTC^0$, an even sharper 
contrast with the classical classes. Indeed, this result implies 
that we can approximate the quantum fast Fourier transform in 
constant depth using fanout. Thus the ``quantum part'' of Shor's 
renowned quantum factoring algorithm can be carried out with a 
quite simple, constant depth quantum circuit that uses the fanout 
operator. 
 
These results suggest the following question: Is fanout really 
necessary to do the quantum Fourier transform in constant depth? 
While so much can be ``reduced" to fan{\it out}, it is far from clear 
how much can be reduced to fan{\it in}\/, even in what appears to 
be its weakest form (i.e., the generalized Toffoli gate). Although 
generalized Toffoli gates can involve just as many bits as fanout 
gates, they may be more feasible to implement and it is 
instructive to investigate their power in constant-depth circuits. 
Note that Cleve and Watrous~\cite{CW} proved that with only one and two 
qubit gates it is not possible to approximate the quantum Fourier 
transform in less 
than log depth, but no similar lower bounds against quantum circuits 
containing gates of unbounded size are known.

 
Our main result, proved in Section~\ref{sec:log-depth}, is that 
one cannot compute parity (and hence fanout) with $\QAC^0$ 
circuits (i.e., in constant depth, {\it without}\/ fanout) using a 
constant number of ancill\ae. This is the first hard evidence
that $\QAC^0$ and $\QAC^0_{wf}$ may be different, and that
fanout may be necessary for all 
the upper bound results mentioned above (it certainly
is if we can get by with only 
constantly many ancill\ae). The issue of the necessity of 
ancill\ae\ in quantum computations is a murky one. 
It is generally accepted that a limited number (polynomially many 
relative to the number of inputs) are needed. This seems 
reasonable as it allows polynomially extra space in which to carry 
out a computation.
However, it is possible to approximate any unitary operator
with a small set of universal gates without ancill\ae\ (although
one apparently needs circuits of great depth and size in
order to do so).
Furthermore, to our knowledge,
no systematic investigation into the absolute
necessity of ancill\ae\ has been done.
They play a crucial role in the present result, in 
which we find the lower bound to be difficult to obtain when more 
than sublinearly many ancill\ae\ are allowed. To help clarify this 
problem, we provide a proof (implicitly claimed, but omitted, in 
Cleve and Watrous) that quantum circuits with gates of bounded
size must be of log depth to 
compute parity (and hence fanout) exactly. In particular, we 
carefully address the problem of including ancill\ae, and show 
that in this case the depth of the circuit must be $\log n$ to 
compute parity, no matter how many ancill\ae\ are used. This is 
given in Section~\ref{sec:one-and-two}. In 
Section~\ref{sec:log-depth}, we allow circuits to
include Toffoli gates of 
unbounded size. It is easiest to see the log-depth lower bound in 
the case of zero ancill\ae, so this result is given first, in 
Theorem~\ref{zero-ancillae}. We then explain how the proof yields 
a depth/ancill\ae\ trade-off, showing that with fewer ancill\ae\ 
one needs greater depth to compute fanout. 
 
We end with some open questions.

\section{Preliminaries}\label{sec:Prelims} 
 
  In this section we set down most of our notational conventions 
and the circuit elements we use. Some acquaintance with 
quantum computational complexity as described in~\cite{nielsenchuang}
or~\cite{KSV} is assumed. 
 
The following notation and terminology 
 will be convenient.  Let $\mc{H}$ denote the 
2-dimensional Hilbert spanned by the computational basis states 
$\ket{0}, \ket{1}$. Let $\mc{H}_1,\ldots,\mc{H}_n$ be $n$ copies of 
$\mc{H}$. By $\mc{B}_{\{1,\ldots,n\}}$ (or simply ``$\mc{B}_n$'' when 
the set notation is clearly understood) we denote the $2^n$-dimensional 
Hilbert space $\mc{H}_1\otimes \cdots \otimes\mc{H}_n$ spanned by the 
usual set of computational basis states of the form 
$\ket{x_1,\ldots,x_n}$, where each $x_i \in \{0,1\}$.  We also 
consider ``quotient spaces of $\mc{B}_{\{1,\ldots,n\}}$ over $m$ 
bits,'' defined as $\mc{B}_{ \{i_1,\ldots,i_m\}} = 
\mc{H}_{i_1}\otimes\cdots\otimes\mc{H}_{i_m}$, where 
$\{i_1,\ldots,i_m\} \subseteq \{1,\ldots,n\}$, which obviously have 
dimension $2^m$.  A ``state over a set of $m$ bits'' is a state in 
such a quotient space.  A quantum gate $G$ corresponds to a unitary 
operator (also denoted $G$) acting on some quotient space 
$\mc{B}_{\{i_1,\ldots,i_m\}}$ of $\mc{B}_n$.  We will say that $G$ 
{\it involves} the bits $i_1,\ldots,i_m$.  We will freely identify 
$G$ with any ``extension by the identity'' that acts on a bigger 
quotient space $\mc{B}_A$ for any set of bits $A \supseteq 
\{i_1,\ldots,i_m\}$, that is, $G$ can be identified with the operator 
$G\otimes I$, where $I$ is the identity on 
$\mc{B}_{A-\{i_1,\ldots,i_m\}}$.  If we fix a state $\ket{\Psi_m}$ 
over $m$ bits $\{i_1,\ldots,i_m\}$, we are effectively restricting 
$\mc{B}_{\{1,\ldots,n\}}$ to the $2^{n-m}$-dimensional linear subspace 
$\ket{\Psi_m}\otimes \mc{B}_{\{1,\ldots,n\} - 
\{i_1,\ldots,i_m\}}$. The space $\mc{B}_{\{1,\ldots,n\} - 
\{i_1,\ldots,i_m\}}$ is referred to as the quotient space of 
$\mc{B}_{\{1,\ldots,n\}}$ {\it complementary} to $\ket{\Psi_m}$. 
 
  A {\it single-qubit gate} is 
a 2$\times$2 unitary matrix (e.g., acting in $\mc{B}_{\{1\}}$). For example, the Hadamard gate $H$ is the 
single-qubit gate, 
\[ H = {1 \over \sqrt{2}} \left[ \begin{array}{cc} 
    1 & 1 \\ 
    1 & -1 \end{array}\right]. 
    \] 
 A {\it generalized Toffoli gate}, which we refer to 
in this paper as simply a {\it Toffoli} gate $T$, transforms computational basis 
states as follows: 
\begin{eqnarray*} 
    T\ket{x_1,...,x_{n}} = \ket{x_1,...,x_{n}, b\oplus \wedge_{i=1}^n x_i} 
\end{eqnarray*} 
A {\it generalized $Z$-gate}, which we refer to as a $Z$-{\it gate} for brevity, 
has the following effect: 
\begin{eqnarray*} 
   Z\ket{x_1,...,x_n} = (-1)^{\bigwedge_{i=1}^n} \ket{x_1,...,x_n}. 
\end{eqnarray*} 
It is not hard to show that, 
$T = HZH$ 
where the Hadamard gate $H$ in this equation is applied to the target 
bit of $T$. Hence we may substitute $Z$-gates for $T$-gates in any circuit 
that allows Hadamards (which will be true throughout the paper). $Z$-gates 
are useful for our purposes since they do not permute computational basis 
states, and thus have no preferred target bit. 
 
  The {\it fanout} gate $F$ and the {\it parity} gate $P$ 
are defined, respectively, by 
\begin{eqnarray*} 
    F\ket{x_1,...,x_n,b} & = & \ket{b\oplus x_1, ..., b\oplus x_n, b},\\ 
   P\ket{x_1,...,x_n,b} & = & \ket{x_1,...,x_n,b\oplus \bigoplus\limits_{i=1}^n x_i}. 
\end{eqnarray*} 
There is no obvious {\it a priori} relation between these operators, 
but as was observed by Moore, $F$ is conjugate to 
$P$ via an $n+1$-fold 
tensor product of Hadamards applied to all the bits: 
\begin{eqnarray}\label{eq:FconjugateP} 
    F = H^{\otimes (n+1)} P H^{\otimes (n+1)} 
\end{eqnarray} 
 
   Recall that Hadamard, phase, CNOT (Toffoli gates for $n=1$), 
and $\pi/8$ gates are a universal 
set of gates in that any unitary operator can be approximated to an 
arbitrary degree of precision with them. Our lower bound techniques 
work against {\it arbitrary} sets of single-qubit gates combined with 
$Z$-gates, which is also a universal set by the above discussion. 
 
  A quantum circuit is constructed out of layers. Each layer $L$ 
is 
a tensor product of a certain fixed set of gates (in our main theorems, 
these will consist of single-qubit and $Z$-gates). A circuit is 
simply a (matrix) product of layers $L_1L_2\cdot\cdot\cdot L_d$. 
(Observe that ``last" layer $L_d$ is actually the one that is 
applied directly to the inputs, and $L_1$ is the output layer.) 
The number of layers $d$ is called the {\it depth} of $C$. 
A circuit $C$ over $n$ qubits is then 
a unitary operator in the $2^n$-dimensional Hilbert space $\mc{B}_{\{1,...,n\}}$. 
Clearly, $C$ computes a unitary 
operator $U$ {\it exactly} if for all computational basis states, 
$C\ket{x_1,...,x_n} = U\ket{x_1,...,x_n}$. This is in general too restrictive, 
however. One must allow for the presence of ``work bits," called 
{\it ancill\ae}, that make extra space available in which to do 
a computation. In that case, in order to exactly compute the operator 
$U$ we extend the Hilbert space in which $C$ acts to the 
$2^{n+m}$-dimensional space spanned by 
computational basis states $\ket{x_1,...,x_n,a_1,...,a_m}$, where 
again $x_i, a_i \in \{0,1\}$, the $a_i$ serving as ancill\ae. 
Then we say that $C$ {\it cleanly computes} $U$ 
if, for any $x_1,...,x_n$ and $y_1,...,y_n$, 
\begin{eqnarray*} 
    \bra{y_1,...,y_n,0,...,0}C\ket{x_1,...,x_n,0,...,0} = 
    \bra{y_1,...,y_n,0,...,0}(U\otimes I)\ket{x_1,...,x_n,0,...,0}, 
\end{eqnarray*}
where $I$ is the identity in the subspace that acts on the
ancill\ae, and the number of 0s in each state above is $m$. 
That is, $C$ does a clean computation if the ancill\ae\ begin 
and end all as 0s. We assume all of our circuits perform clean 
computations. This is a reasonable constraint, since only then is it easy 
to compose the circuits. 
 
  Lastly, all circuits should be understood to be elements of 
an infinite {\it family} of circuits $\{C_n | n \ge 0\}$, where 
$C_n$ is a quantum circuit for $n$ qubits. 
 
\section{Fanout Requires Log Depth with Bounded Size 
Gates}\label{sec:one-and-two} 

It is easy to see that, by an obvious divide-and-conquer strategy,
we can compute parity in depth $\log n$ using just CNOT gates
and 0 ancillae.
In this section we prove this is optimal for any bounded size
multi-qubit gates, and furthermore that no number of ancillae
help to reduce the depth of the circuit.
 
Let 
$C = L_1\cdot\cdot\cdot L_d$ consist entirely of arbitrary two-qubit gates 
and 
single-qubit gates. (The extension to arbitrary, but fixed, size gates 
is straightforward.) Further suppose that 
$M$ is an observable on a single qubit in the 
last layer. Let $L'_1$ denote the gate whose 
output $M$ is measuring. $L'_1$ could be a two-qubit 
or a single-qubit gate. In either case, $L_1 
= L'_1 \otimes R_1$, where $R_1$ is the tensor 
product of all the other gates 
in that layer, if any. 
More generally, we decompose 
layer $i$ similarly, writing $L_i = L'_i \otimes R_i$, 
where $L'_i$ is a transformation that acts on some
subset of the bits, and $R_i$ acts on the rest. 
 
\begin{Lemma}\label{one-and-two} 
 For each $d$, there are layers $L'_1,...,L'_d$ such that 
$$ L_d^{\dagger}L_{d-1}^{\dagger}\cdot\cdot\cdot L_1^{\dagger} 
   M L_1\cdot\cdot\cdot L_{d-1}L_d = 
   {L'}_d^{\dagger}{L'}_{d-1}^{\dagger}\cdot\cdot\cdot {L'}_1^{\dagger} 
   M {L'}_1\cdot\cdot\cdot {L'}_{d-1}{L'}_d$$ 
where, for each $i$, $L'_i$ acts on at most $2^i$ bits. Furthermore, 
for each $i$, ${L'}_i$ acts on bits with indices in some set $S_i$ such 
that 
$S_d \supseteq S_{d-1} \supseteq ...\supseteq S_1$. 
\end{Lemma} 
 
\begin{figure}
\begin{center}
\begin{picture}(0,0)%
\includegraphics{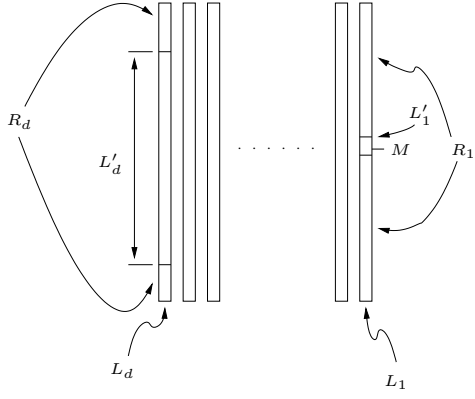}%
\end{picture}%
\setlength{\unitlength}{2013sp}%
\begingroup\makeatletter\ifx\SetFigFont\undefined%
\gdef\SetFigFont#1#2#3#4#5{%
  \reset@font\fontsize{#1}{#2pt}%
  \fontfamily{#3}\fontseries{#4}\fontshape{#5}%
  \selectfont}%
\fi\endgroup%
\begin{picture}(5967,4795)(751,-5969)
\put(1876,-3211){\makebox(0,0)[lb]{\smash{{\SetFigFont{6}{7.2}{\rmdefault}{\mddefault}{\updefault}$L'_d$}}}}
\put(2026,-5761){\makebox(0,0)[lb]{\smash{{\SetFigFont{6}{7.2}{\rmdefault}{\mddefault}{\updefault}$L_d$}}}}
\put(5401,-5911){\makebox(0,0)[lb]{\smash{{\SetFigFont{6}{7.2}{\rmdefault}{\mddefault}{\updefault}$L_1$}}}}
\put(6226,-3061){\makebox(0,0)[lb]{\smash{{\SetFigFont{6}{7.2}{\rmdefault}{\mddefault}{\updefault}$R_1$}}}}
\put(5476,-3061){\makebox(0,0)[lb]{\smash{{\SetFigFont{6}{7.2}{\rmdefault}{\mddefault}{\updefault}$M$}}}}
\put(751,-2686){\makebox(0,0)[lb]{\smash{{\SetFigFont{6}{7.2}{\rmdefault}{\mddefault}{\updefault}$R_d$}}}}
\put(5701,-2611){\makebox(0,0)[lb]{\smash{{\SetFigFont{6}{7.2}{\rmdefault}{\mddefault}{\updefault}$L'_1$}}}}
\end{picture}%
 \caption{Decomposition of the layers of the circuit 
$C$.}\label{fig:layers_1} 
\end{center}
\end{figure}

Figure~\ref{fig:layers_1} makes the notation a little 
clearer. Note that the {\it input} will, as usual, be 
on the {\it left}, but it doesn't enter the claim (or 
the following argument) at all.\\ 
 
\begin{proof} 
The proof of Lemma~\ref{one-and-two} is by induction on $d$. First consider $d=1$. 
Then consider the operator $L_1^{\dagger}M L_1$. By the 
observations above, we may write $L_1 = L'_1 \otimes R_1$, 
where $L'_1$ is either a single or two-qubit gate. 
So, 
$$L_1^{\dagger}M L_1 = (L'^{\dagger}_1\otimes R^{\dagger}_1)M (L'_1 \otimes R_1) 
= L'^{\dagger}_1 M L'_1,$$ 
by virtue of the fact that $M$ and $R_1$ commute. Since ${L'}_1$ only 
depends on $ \le 2$ qubits, this establishes the result for $d =1$. 
 
Now suppose that we can write, 
$$ L_d^{\dagger}L_{d-1}^{\dagger}\cdot\cdot\cdot L_1^{\dagger} 
   M L_1\cdot\cdot\cdot L_{d-1}L_d = 
   {L'}_d^{\dagger}{L'}_{d-1}^{\dagger}\cdot\cdot\cdot {L'}_1^{\dagger} 
   M {L'}_1\cdot\cdot\cdot {L'}_{d-1}{L'}_d$$ 
where, for each $i$, $L'_i$ acts on at most $2^i$ bits. In particular, 
note that $L'_d$ acts on at most $2^d$ bits. Suppose that 
$L'_d$ acts on indices in the set $S_d$ (where $S_d$ has size $ \le 2^d$). 
Now by the induction hypothesis, 
$$ L_{d+1}^{\dagger}L_{d}^{\dagger}\cdot\cdot\cdot L_1^{\dagger} 
   M L_1\cdot\cdot\cdot L_{d}L_{d+1} = 
   L_{d+1}^{\dagger}{L'}_{d}^{\dagger}\cdot\cdot\cdot {L'}_1^{\dagger} 
   M {L'}_1\cdot\cdot\cdot {L'}_{d}L_{d+1},$$ 
and $S_d \supseteq S_{d-1} \supseteq ...\supseteq S_1$. 
 
The gates in $L'_d$ involve at most the bits in $S_d$. Since the 
circuit only contains at most two-qubit gates, all the gates in 
$L_{d+1}$ involving bits in $S_d$ can act on at most $2^{d+1}$ 
bits. Let the tensor product of these gates be denoted by $L'_{d+1}$, 
and $S_{d+1}$ denote the set of bits on which $L'_{d+1}$ acts. 
Clearly $S_{d+1} \supseteq S_d$. Then for some tensor product 
of single and two-qubit gates $R_{d+1}$ we may write $L_{d+1} 
 = L'_{d+1}\otimes R_{d+1}$. 
Since $R_{d+1}$ acts 
on bits not in $S_{d+1}$, it commutes with all the ${L'}_i$ and $M$, 
which only act on bits inside $S_{d+1}$. Hence $R_{d+1}$ ``cancels out'' 
and we have the desired relation. 
\end{proof} 
 
\begin{Theorem}\label{one-and-two-theorem} 
Let $C$ be a quantum circuit on $n$ inputs 
of depth $d$, consisting of single-qubit and two-qubit gates, 
with any number of ancill\ae\ that 
cleanly computes parity exactly. Then $d \ge \log n$. 
If $C$ computes fanout in the same way, then $d \ge \log n - 2$. 
\end{Theorem} 
\begin{proof} 
Let $C = L_1\cdot\cdot\cdot L_d$ as in Lemma~\ref{one-and-two}. Suppose 
$C$ uses $m$ ancill\ae, and that it cleanly computes the parity operator $P$ 
in depth $d < \log n$. 
It follows that 
for any $x_1,...,x_n, b$ and any measurement operator $M$ on the 
target bit, 
\begin{eqnarray}\label{eq:CMC=PMP} 
   \bra{x_1,...,x_n,b,0,...,0} C^{\dagger} M C\ket{x_1,...,x_b,0,...,0} 
   = \bra{x_1,...,x_n,b} P M P\ket{x_1,...,x_n,b}. 
\end{eqnarray} 
By Lemma~\ref{one-and-two}, 
\begin{eqnarray*} 
    C^{\dagger}M C = 
      L_d^{\dagger}L_{d-1}^{\dagger}\cdot\cdot\cdot L_1^{\dagger} 
   M L_1\cdot\cdot\cdot L_{d-1}L_d  = 
     {L'}_d^{\dagger}{L'}_{d-1}^{\dagger}\cdot\cdot\cdot {L'}_1^{\dagger} 
   M {L'}_1\cdot\cdot\cdot {L'}_{d-1}{L'}_d, 
\end{eqnarray*} 
where the operator ${L'}_1\cdot\cdot\cdot{L'}_d$ acts on at most $2^d$ inputs. 
Since $2^d < n$, there is an input on which that operator does not act. 
Hence the value on the left hand side of eq.~(\ref{eq:CMC=PMP}) 
remains unchanged if we can flip some 
$x_i$. However, the outcome 
of the measurement on the parity gate on the right hand side depends on 
every input, which is a contradiction. 
 
  The second assertion in the Theorem follows from eq.~(\ref{eq:FconjugateP}). 
\end{proof} 
 
It is clear that if we have a family of circuits that use a fixed set 
of multi-qubit gates with arity independent of $n$, that a similar proof 
will work. Thus we have the following as a corollary of the proof of 
Theorem~\ref{one-and-two-theorem}: 
 
\begin{Cor}\label{one-and-two-corollary} 
Let $C$ be a quantum circuit on $n$ inputs 
of depth $d$, consisting of single-qubit and multi-qubit gates of size 
$O(1)$, 
with any number of ancill\ae, that 
cleanly computes parity, or fanout, exactly. Then $d = \Omega(\log n)$. 
\end{Cor} 
 
\section{Parity Requires Log Depth with Few Ancill\ae}\label{sec:log-depth} 
 
  In this section we treat circuits that contain Toffoli gates or, 
equivalently, $Z$-gates, 
of arbitrary size (i.e., that can depend on $n$). The technique of 
the preceding section does not work in this case. This is because 
the large gates in general do not cancel, since they may not commute 
with the measurement operator $M$. 
 
  To see how to proceed, it is useful to briefly 
consider classical circuits with similar constraints. 
Suppose we have a classical circuit with NOT gates and 
unbounded fan-in AND and OR gates, 
but that we do {\it not} allow any fanout. Once inputs (or outputs 
of other gates) are used in either an AND or an OR gate, they 
can not be used again. It is obvious that if such a circuit has 
constant depth, it cannot 
compute such functions as parity. The AND and OR gates can be killed 
off by restricting a small set of inputs, resulting in a constant 
function, while parity depends on all the inputs. 
 
  In the quantum case, it appears again that the only thing to 
do is to attempt to ``kill off" the large Toffoli gates. However, 
the quantum case is much more subtle since we must face the 
fact that intermediate states are a superposition of computational 
basis states, and furthermore that 
the $Z$-gates, in combination with the 
single-qubit gates, may cause entanglement. 
 
As before, write $C = L_1L_2\cdots L_d$. Thus 
the circuit $C$ transforms the state $\ket{\Psi}$ to $L_1\cdots L_d 
\ket{\Psi}.$ We assume wlog that each layer $L_i$ is a tensor product 
of $Z$-gates and single-qubit gates. 
Further assume wlog that a specific bit 
(say, the $n^{th}$ bit) of $C$ serves as the output or {\it target} 
bit (which eventually is supposed to agree with the output bit of a 
parity gate). 
 
Our main technical lemma is easiest to see in the case that $C$ has 
{\it no ancill\ae}, which we assume until later in the section: 
 
\begin{Lemma}\label{lem:main} 
 Let $C$ be a circuit as described above, with no ancill\ae. 
 Then for each $1 \le k \le 
 d$, there exists a state $\ket{\Psi_k}$ over at most $2^k$ bits 
 such that for any state $\ket{R}$ in the quotient space of $\mc{B}_n$ 
 complementary to $\ket{\Psi_k}$, the state $L_1L_2\cdots L_k (\ket{R} 
 \otimes \ket{\Psi_k})$ has a 0 in the target position of $C$. 
\end{Lemma} 
\begin{proof} 
The proof is by induction on $k$. First let $k=1$. There are two 
cases: 
\begin{enumerate} 
\item 
In layer $L_1$, the target is the output of a single-qubit gate 
$S$. Then let the state $\ket{\Psi_1} = S^{\dagger} \ket{0}$ over the 
$n^{th}$ bit.  Now we may write $L_1 = L_1' \otimes S$, where $L_1'$ 
acts on the quotient space $\mc{R}$ complementary to 
$\ket{\Psi_1}$. No matter what state $\ket{R} \in \mc{R}$ we choose 
over the bits $\{1,\ldots,n-1\}$, it follows that $L_1(\ket{R}\otimes 
\ket{\Psi_1}) = (L_1'\ket{R})\otimes (S\ket{\Psi_1}) = (L_1'\ket{R}) 
\otimes \ket{0}$ has a 0 in the $n^{th}$ position. 
\item 
In layer $L_1$, the target is the output of a $Z$-gate.  Write $L_1 = 
L_1'\otimes G$, where $G$ is this $Z$ gate.  In this case, we choose 
$\ket{\Psi_1} = \ket{0}$ over the $n^{th}$ bit.  Now $G$ acts both on 
$\ket{\Psi_1}$ as well as the complementary quotient space $\mc{R}$ 
(via extension by the identity).  But since $G$ involves a bit that is 
0 (i.e., the $n^{th}$ bit), $G$ is equivalent to the unit matrix in 
$\mc{R}$.  Hence for any state $\ket{R}\in \mc{R}$, 
$L_1(\ket{R}\otimes\ket{\Psi_1}) = (L_1'\otimes G) 
(\ket{R}\otimes\ket{\Psi_1}) = (L_1'\ket{R})\otimes\ket{0}$ again has 
a 0 in the $n^{th}$ position. 
\end{enumerate} 
 
Now suppose the assertion is true for $k-1$ where $k > 1$.  We will 
show that it remains true for $k$.  Suppose the $\ket{\Psi_{k-1}}$ in 
the assertion is a state over the (at most) $2^{k-1}$ bits in the set 
$K_{k-1}$.  Let $R_{k-1}$ denote the rest of the bits $\{1,\ldots,n\} 
- K_{k-1}$.  Thus $\ket{\Psi_{k-1}}$ is a state in $\mc{B}_{K_{k-1}}$, 
and the quotient space complementary to $\ket{\Psi_{k-1}}$ is 
$\mc{B}_{R_{k-1}}$, which for convenience we denote by 
$\mc{R}_{k-1}$. We specify the state $\ket{\Psi_k}$ as follows: Start 
with $K_k := K_{k-1}$ and $R_k := R_{k-1}$. If a $Z$-gate $G$ in $L_k$ 
involves bits both in $K_k$ and in $R_k$, we remove a \emph{single} 
bit from $R_k$ on which $G$ acts, add it to $K_k$, declare the gate 
$G$ \emph{killed}, and remove $G$ from further consideration. 
Continue until all such $Z$-gates have been killed.  Since each bit in 
$K_{k-1}$ can be involved with at most one $Z$-gate in $L_k$, the 
number of bits added to $K_k$ (and removed from $R_k$) in this process 
is at most $2^{k-1}$.  Let $L^{(K)}_k$ denote the gates in $L_k$ that 
involve the bits in $K_k$, excluding the $Z$-gates that have been 
killed.  Then finally, we define the state $\ket{\Psi_k}$ as the 
tensor product of $L^{(K)\dagger}_k\ket{\Psi_{k-1}}$ with the state in 
which all the bits in $K_k - K_{k-1}$ are 0. 
 
Note that $\ket{\Psi_k}$ is a state over at most $2\cdot 2^{k-1} = 
2^k$ bits, as seen in Figure~\ref{fig:layers}. 

\begin{figure}
\begin{center}
\begin{picture}(0,0)%
\includegraphics{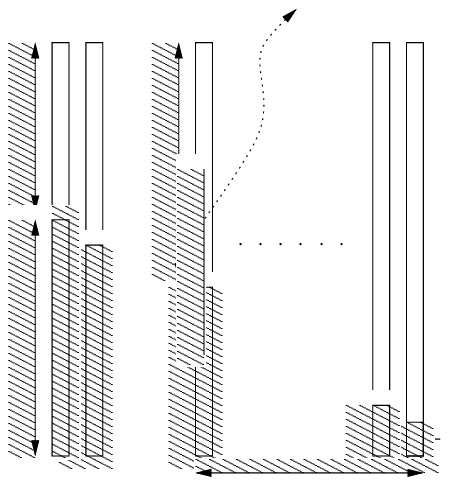}%
\end{picture}%
\setlength{\unitlength}{2131sp}%
\begingroup\makeatletter\ifx\SetFigFont\undefined%
\gdef\SetFigFont#1#2#3#4#5{%
  \reset@font\fontsize{#1}{#2pt}%
  \fontfamily{#3}\fontseries{#4}\fontshape{#5}%
  \selectfont}%
\fi\endgroup%
\begin{picture}(4655,4606)(1097,-5360)
\put(4126,-886){\makebox(0,0)[lb]{\smash{{\SetFigFont{6}{7.2}{\rmdefault}{\mddefault}{\updefault}$Z$}}}}
\put(1351,-2161){\makebox(0,0)[b]{\smash{{\SetFigFont{6}{7.2}{\rmdefault}{\mddefault}{\updefault}$R_d$}}}}
\put(1351,-3961){\makebox(0,0)[b]{\smash{{\SetFigFont{6}{7.2}{\rmdefault}{\mddefault}{\updefault}$K_d$}}}}
\put(3976,-5311){\makebox(0,0)[lb]{\smash{{\SetFigFont{6}{7.2}{\familydefault}{\mddefault}{\updefault}$k$}}}}
\put(5476,-4786){\makebox(0,0)[lb]{\smash{{\SetFigFont{6}{7.2}{\familydefault}{\mddefault}{\updefault}$0$}}}}
\put(2701,-2236){\makebox(0,0)[lb]{\smash{{\SetFigFont{6}{7.2}{\familydefault}{\mddefault}{\updefault}$R_k$}}}}
\put(2701,-4111){\makebox(0,0)[lb]{\smash{{\SetFigFont{6}{7.2}{\familydefault}{\mddefault}{\updefault}$K_k$}}}}
\put(2551,-2986){\makebox(0,0)[lb]{\smash{{\SetFigFont{6}{7.2}{\familydefault}{\mddefault}{\updefault}. . .}}}}
\end{picture}%
\caption{The sets $K_k$ and $R_k$. A $Z$ gate that involves bits 
in both sets is shown.}\label{fig:layers} 
\end{center}
\end{figure}

Let $\mc{R}_k$ denote the quotient space complementary 
to $\ket{\Psi_k}$.  Clearly, $\mc{R}_k = \mc{B}_{R_k}$. Now let 
$\ket{R}$ be any state in $\mc{R}_k$ (equivalently, over the bits in 
$R_k$), and apply $L_k$ to $\ket{R}\otimes \ket{\Psi_k}$. Let 
$L^{(R)}_k$ denote the gates in $L_k$ acting in $\mc{R}_k$, again 
excluding the $Z$-gates that have been killed.  Note that any $Z$-gate 
in layer $L_k$ that involves bits in $K_k$ as well as $R_k$ acts as 
the identity on $\mc{R}_k \otimes \ket{\Psi_k}$, by the construction 
of $\ket{\Psi_k}$.  Thus we have eliminated these gates from $L_k$ 
without any loss of generality.  Thus, $$L_k(\ket{R}\otimes 
\ket{\Psi_k}) = (L^{(R)}_k \otimes L^{(K)}_k) (\ket{R}\otimes 
\ket{\Psi_k}) = (L^{(R)}_k \ket{R})\otimes (L^{(K)}_k \ket{\Psi_k}).$$ 
Now $L^{(K)}_k\ket{\Psi_k}$ is the tensor product of 
$\ket{\Psi_{k-1}}$ with a number of $\ket{0}$ states. So we conclude 
that $L_k (\ket{R} \otimes \ket{\Psi_k})$ is of the form 
$\ket{R'}\otimes \ket{\Psi_{k-1}} $ for some state $\ket{R'} \in 
\mc{R}_{k-1}$. Then, 
\begin{eqnarray*} 
  L_1L_2\cdots L_{k-1}L_k(\ket{R}\otimes\ket{\Psi_k}) 
  = L_1L_2\cdots 
  L_{k-1}(\ket{R'}\otimes\ket{\Psi_{k-1}}). 
\end{eqnarray*} 
By the induction hypothesis, the right hand side of the above 
equation has a 0 target bit, which proves the lemma. 
\end{proof} 
 
\paragraph{Remark.} 
With a bit more careful analysis, Lemma~\ref{lem:main} can be improved 
to the following: 
 
\begin{Lemma}\label{lem:main-better} 
 Let $C$ be a circuit as described above. Then for each $1 \le k \le 
 d$, there exists a state $\ket{\Psi_k}$ over at most $2^{k/2}$ bits 
 such that for any state $\ket{R}$ in the quotient space of $\mc{B}_n$ 
 complementary to $\ket{\Psi_k}$, the state $L_1L_2\cdots L_k (\ket{R} 
 \otimes \ket{\Psi_k})$ has a 0 in the target position of $C$. 
\end{Lemma} 
 
The difference is that now $\ket{\Psi_k}$ is over only $2^{k/2}$ bits 
instead of $2^k$.  Instead of giving a formal proof, we will just 
sketch the reasons for Lemma~\ref{lem:main-better}.  When some bit 
(the $i^{th}$ bit, say) is moved from $R_k$ to $K_k$, it is set to the 
$\ket{0}$ state.  Consider the gate $G$ (if any) in $L_{k+1}$ that 
involves this bit.  If $G$ is a single-qubit gate, then no $Z$-gate is 
killed involving the $i^{th}$ bit, so no additional bit needs to be 
added to $K_{k+1}$ for the sake of the $i^{th}$ bit.  If $G$ is a $Z$-gate, 
then the $i^{th}$ bit alone is enough to kill $G$, since this bit is 
already 0.  So again, no additional bit must be added to $K_{k+1}$ to 
kill $G$.  Thus $k$ must increase by $2$ for the size of $K_k$ to 
double.  Note that we handled the base case of Lemma~\ref{lem:main} 
this way, obtaining a state over $1 = 2^0$ bits. 
 
 
\begin{Theorem}\label{zero-ancillae} 
Let $C$ be a circuit of depth $d$ consisting of single-qubit gates 
and $Z$-gates, and uses 0 ancill\ae.  If $d 
< 2\log n$, then $C$ cannot compute $P$. 
\end{Theorem} 
\begin{proof} 
Suppose $C = P$.  Then for any input state, the target bit of $C$ is 0 
iff the target bit of $P$ is 0.  By Lemma~\ref{lem:main-better}, 
there exists a state $\ket{\Psi}$ on at most $2^{d/2} < n$ bits such 
that, for any state $\ket{R}$ on the remaining $n - 2^{d/2}$ 
bits, $C(\ket{R}\otimes\ket{\Psi})$ has a 0 value for the 
target. First let $\ket{R}$ be the state with 0s in all $n-2^{d/2}$ 
positions (since $n - 2^{d/2} > 0$, such positions exist).  Then 
$P(\ket{R}\otimes\ket{\Psi})$ has a 0 target. This is only possible if 
the state $\ket{\Psi}$ is in a quotient space of $\mc{B}_n$ spanned by 
computational basis states in which an even number of the variables 
are 1.  Now change one of the bits of $\ket{R}$ from 0 to 1. The target 
of $C(\ket{R}\otimes\ket{\Psi})$ still has the value 0, but the target 
of $P(\ket{R}\otimes\ket{\Psi})$ must change to 1, which contradicts 
the assumption that $C = P$. 
\end{proof} 
 
Since fanout and parity are equivalent up to depth 3 (with 0 
ancill\ae), we have immediately the following. 
 
\begin{Cor} 
 Let $C$ be a circuit of depth $d$ consisting of single-qubit gates 
and $Z$-gates, and uses 0 ancill\ae. Then, if $d < 
 2\log n - 2$, $C$ cannot compute the fanout operation. 
\end{Cor} 
 
We now consider the case in which our circuit has a non-zero number 
of ancill\ae. 
Firstly, it is clear that Lemmas~\ref{lem:main} 
and~\ref{lem:main-better} work if we set a target and all ancill\ae\ 
to 0 at the same time.  If there are $a$ many ancill\ae, then we are 
setting $a+1$ ``outputs.'' The conclusion of the analogous Lemma for 
$a$ ancill\ae\ would then be that the state $\ket{\Psi}$ is over 
$(a+1)2^{d/2}$ bits (since the number of ``committed'' bits doubles 
with each second layer, as in Lemma~\ref{lem:main-better}).  These 
bits may include all the ancill\ae, and assuming that $C$ does a clean 
computation, $\ket{\Psi}$ will be 0 on the ancill\ae\ (since they 
must all start out as 0 in order to return to their final 
value of 0).  Therefore, if $n > (a+1)2^{d/2}$, the state $\ket{R}$ 
does not involve any of the ancill\ae\ and is thus free to take on any value. 
Thus if $n > (a+1)2^{d/2}$, the output of $C$ is insensitive to 
changes in at least one of the inputs, and hence the circuit is 
defeated as before.  Note we have a depth/ancill\ae\ trade-off as a 
result.  We thus have the following corollary of the proof of 
Theorem~\ref{zero-ancillae}: 
 
\begin{Cor} 
 Let $C$ be a circuit of depth $d$ consisting of single-qubit gates 
and $Z$-gates. Then, if $C$ cleanly computes the parity function 
with $a$ ancill\ae, then $d \ge 2\log (n/(a+1))$. 
\end{Cor} 
 
We conjecture that $d$ must be at least $2\log n$ 
no matter what $a$ is. 

 

We offer an alternative interpretation of our result that arose
out of conversations with L.~Longpr{\'e}. Let us say that a quantum circuit
$C$
{\it robustly computes} a unitary operator $U$ if $C$ computes $U$
cleanly and, in addition, if its output is insensitive to the inititial
state of the ancill\ae. Thus the ancill\ae\ of $C$ can start out in
any state whatsoever; the circuit $C$ is guaranteed to return the
ancill\ae\ to that state in the end, and always gives the same
answer. This of course puts a much
stronger constraint on the circuit (since in the usual model we only
insist on a clean computation when the ancill\ae\ are initialized to
0), but such circuits can be useful
 (e.g., see exercise 8.5 in Kitaev
et al.~\cite{KSV}). It is not hard to see that in this case, if $C$ consists
only of single-qubit and Toffoli gates, then it must have depth
$\log n$ to compute
parity, regardless of the number of ancill\ae.

\section{Conclusions and Open Problems}

Following the line of earlier work of Green et al., H{\o}yer and Spalek,
and Cleve and Watrous \cite{CW},
our main result gives an optimal, O(log n) lower bound on the depth of
$\QAC$-type
circuits computing fanout, in the presence of limited (slightly 
sublinear)
 numbers of ancil\ae. 
It would clearly be desirable to extend our result to obtain the same
conclusion when polynomially many (or an unlimited number of)
ancill\ae\ are allowed, and
thus to prove that $\QAC^0 \not = \QAC^0_{wf}$.

The role of ancill\ae\ in quantum computation has not
received much detailed attention. Prompted by our considerations here,
there are several interesting
questions that arise. One issue is the necessity of ancillae for specific 
quantum computations or classes of quantum computations. Is there a problem 
that can be done in constant
depth with ancill\ae\ but which requires $\log n$ depth without ancill\ae?
Similarly, are there computational problems for which $\log n$ depth is possible
with ancill\ae\ but without ancill\ae, polynomial depth is needed? 
In general, how many ancill\ae\ are needed for specific problems? 
Is there a general 
tradeoff that can be proved between numbers of ancill\ae\ and circuit depth?

While much has recently been learned concerning constant depth
circuit classes, a few interesting 
questions still remain. 
It would be worthwhile to be able to distinguish between the power of
quantum gates of unbounded arity. We have seen that Toffoli and 
Z gates (which are equivalent up to constant depth)
are weaker than parity and fanout 
(which are equivalent not only to each other but also, for all intents
and purposes, to other mod gates,
 threshold gates and the quantum Fourier transform).
Are there other
natural types of gates that lie between these 
two classes, or is every gate either equivalent, up to constant depth,
to either  single qubit and CNOT gates, or to Toffoli gates, or to
parity? It would also be of interest to characterize exactly
what can be computed in constant depth using only single qubit and CNOT gates,
as even very optimistically, this is the kind of circuit that might
be built in the not too distant future.

\section{Acknowledgements}
We thank Luc Longpr{\'e} for helpful discussions and comments on this paper.
This work was supported in part by the National Security Agency (NSA)
and Advanced Research and Development Agency (ARDA) under Army Research
Office (ARO) contract numbers DAAD~19-02-1-0058 (for M.~Fang, S.~Homer,
and F.~Green) and DAAD~19-02-1-0048 (for S.~Fenner and Y.~Zhang).


\begin{thebibliography}{1}

\bibitem{CW} R.~Cleve and J.~Watrous,
``Fast parallel circuits for the quantum Fourier transform,''
{\em Proceedings of the 41st Annual Symposium on Foundations
of Computer Science} (2000),  526--536.

\bibitem{GHMP} F.~Green, S.~Homer, C.~Moore and C.~Pollett,
"Counting, Fanout and the Complexity of Quantum ACC,"
{\em Quantum Information and Computation} {\bf 2} (2002)
35--65.

\bibitem{HS} P.~H{\o}yer and R.~Spalek,
``Quantum circuits with unbounded fan-out,''
{\em 20th STACS Conference, 2003}, LNCS 2607, 234--246.

\bibitem{KSV} A.~Yu.~Kitaev, A.~H.~Shen, and M.~N.~Vyalyi,
Classical and Quantum Computation, American Mathematical Society,
2002.

\bibitem{moore99}
Cristopher Moore.
\newblock Quantum Circuits: Fanout, Parity, and Counting.
\newblock In Los Alamos Preprint archives (1999), quant-ph/9903046.

\bibitem{nielsenchuang}
M.~A.~Nielsen and I.~L.~Chuang, Quantum Computation and Quantum Information,
Cambridge University Press, 2001.

\bibitem{Razborov}
A.~A.~Razborov, Lower bounds on the size of bounded
depth networks over a complete basis with logical addition, {\em
Matematicheskie Zametki} {\bf 41} (1987) 598-607. English translation in {\em
Mathematical Notes of the Academy of Sciences of the USSR} {\bf 41} (1987)
333-338.

\bibitem{Smolensky}R.~Smolensky, Algebraic methods in the
theory of lower bounds for Boolean circuit complexity, in {\em
Proceedings of the 19th Annual ACM Symposium on Theory of
Computing} (1987) 77-82.

\end{thebibliography}
\end{document}